# BEST-2 EXPERIMENT WITH $^{58}$Co NEUTRINO SOURCE

V.N.Gavrin [a], V.V. Gorbachev [a*], T.V.Ibragimova [a], V.A.Matveev [b]

[a] *Institute for Nuclear Research, Russian Academy of Sciences, Moscow*

[b] *Joint Institute for Nuclear Research, Dubna*

*e-mail: vvgor_gfb1@mail.ru

**Abstract** - The article describes a new experiment with an artificial neutrino source $^{58}$Co on a gallium target GGNT (SAGE). The goal of the experiment is to study the gallium anomaly. The experiment makes it possible to find the parameters of oscillation transitions of electron neutrinos to sterile states in a wide range of parameters. Including the parameter $\Delta m^2$, the experimental determination of which usually causes significant difficulties. An important feature of the experiment is the possibility of identifying the dependence of the gallium anomaly on the neutrino energy.

**1. Introduction**

In recent years, attempts have been made to explain the unusual results of a number of neutrino experiments – LSND [1], MiniBooNE [2-4], short-baseline reactor experiments [5-7], gallium experiments with artificial neutrino sources [8-11] – by the features of individual experiments, in which the systematics has not been sufficiently studied, and by general reasons associated with physical phenomena that affect the measurements of all such different detectors at once [12]. One attractive explanation is the sterile neutrino hypothesis, which assumes the existence of a fourth neutrino flavor that does not have a charged partner in the lepton sector and does not interact with matter weakly [13]. Such neutrinos are produced only in oscillations of neutrinos of ordinary flavors and manifest themselves either through the suppression of interactions of ordinary neutrinos, as in the examples of the reactor and gallium anomalies, or through the appearance of neutrinos of other flavors under conditions where they are forbidden by oscillations of the three-neutrino model, as in LSND and MiniBooNE .

In recent years, several experiments have been conducted to test the sterile neutrino hypothesis, yielding contradictory results. For example, in the experiments with reactor antineutrinos STEREO [14,15], PROSPECT [16], DANSS [17], NEOS [18], and in the experiment to measure the mass of the electron neutrino KATRIN [19], the results are well described within the framework of the 3-flavor neutrino scheme. There are also no indications of sterile neutrinos in experiments to search for neutrinoless beta decay [20].

Data from the T2K accelerator experiment [21] exclude part of the oscillation parameter space and simultaneously highlight the region of allowed oscillation parameters in the hypothesis of one sterile neutrino.

In the Neutrino-4 reactor experiment [22], not only was a significant suppression of the antineutrino counting rate obtained relative to the expected value obtained, but the oscillation parameters in the sterile neutrino hypothesis were also found.

Also, a recent work by the IceCube collaboration has provided evidence for the existence of sterile neutrinos and found the parameters of oscillatory transitions between sterile neutrinos and muon and tau neutrinos [23].

---

In the BEST experiment [24,25] with the gallium target of the SAGE solar neutrino experiment, irradiated with neutrinos from an artificial $^{51}$Cr source, confirmation of the gallium anomaly was obtained at the level of 4 standard deviations (4σ), and under some additional assumptions 5σ [26].

Thus, at present the hypothesis of sterile neutrinos remains relevant. Theoretical limitations on the existence of sterile neutrinos are mainly related to limitations on the mass of the fourth mass state [27], while direct experiments cannot confidently confirm or refute this hypothesis.

In this paper, we propose the BEST-2 experiment, which implements a scheme for recording just the oscillation nature of the counting rate variation of neutrino interactions in the detector depending on the distance for a certain interval of values of the oscillation parameter $\Delta m^2$. If the sterile neutrino hypothesis is correct, and the oscillation parameters are in the detector sensitivity range, the experiment will not only find such oscillations, but also determine their parameters.

The experiment will investigate the dependence of the gallium anomaly not only on the distance from the source to the interaction point, but also on the neutrino energy.

**2. Gallium anomaly**

The efficiency of the solar neutrino experiments SAGE and GALLEX [8-11] was tested in calibration experiments with intense artificial sources of $^{51}$Cr and $^{37}$Ar. The rates of registration of monochromatic neutrinos from sources obtained in four experiments turned out to be lower than expected by 2.6σ [12]: $R = \dfrac{v_{measured}}{v_{predicted}} = 0.87 \pm 0.05$.

A detailed examination of each procedure independently in each experiment showed that systematic errors could not explain the low measured count rate.

The result found was called the gallium anomaly, the explanation of which began to be associated with "new physics", i.e. the reasons must be sought in the unknown properties of neutrinos.

To study the gallium anomaly, in 2019, on the basis of the SAGE experiment, the BEST experiment was conducted with a $^{51}$Cr source with an activity of 3.4 MCi, approximately 7 times greater than in previous source experiments [25]. In the experiment, the gallium target (48 t of metallic gallium) was divided into two parts located at different distances from the source. The source was placed in the center of both target zones, located one inside the other - a spherical zone inside a cylindrical zone. With equal average gallium thicknesses in both zones, the expected counting rates from the source in the absence of "new physics" effects were also the same. Dividing the target made sensitivity of the experiment to the dependence of the counting rate on the distance traveled by the neutrino.

The BEST experiment also yielded a lower counting rate than expected, confirming the gallium anomaly at the 4σ level: $R = 0.80 \pm 0.05$. Figure 1 [28] shows the counting rates in two target zones as functions of the $\Delta m^2$ parameter. The experiment was sensitive to two values of this parameter, at which the difference in counting rates in the two zones was maximal – $\Delta m^2 = 0.9$ eV$^2$ and 1.8 eV$^2$. The counting rates found in the measurements were identical within the error, so it was not possible to find or limit the value of the $\Delta m^2$ parameter in the BEST experiment.

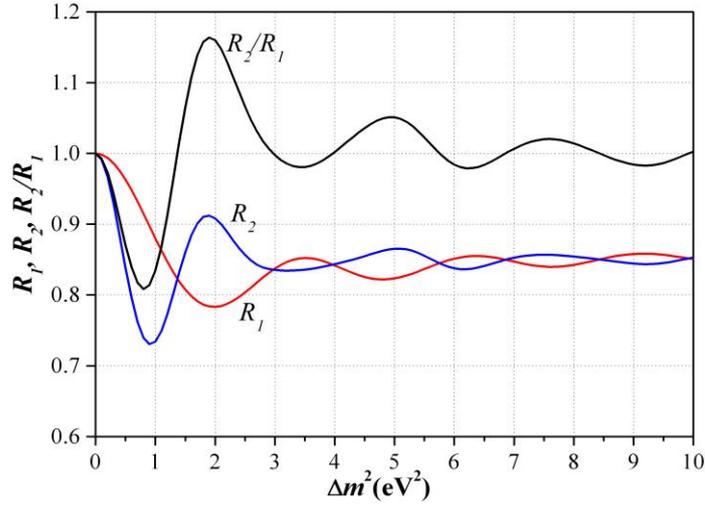

Fig. 1. Dependences of expected counting rates in two zones of the gallium target in the BEST experiment and their ratio on the parameter $\Delta m^2$ for a fixed value of the parameter $\sin^2 2\theta = 0.30$

### 3. The purpose of the experiment BEST-2

BEST-2 experiment with the neutrino source $^{58}$Co is designed for a detailed study of the gallium anomaly. The experiment will obtain data on the dependence of the gallium anomaly on the neutrino energy $E$ and on the distance between the points of neutrino birth and capture $L$.

The leading hypothesis explaining the gallium anomaly today is that neutrinos oscillate at short distances ($\Delta m^2 \sim 1$ eV$^2$). Oscillations of electron neutrinos with energy $E$ at a distance $L$ from the source are determined through the probability of survival of electron neutrinos:

$$P_{ee} = 1 - \sin^2 2\theta \cdot \sin^2 \frac{1.27 \Delta m^2 L}{E}, \qquad (1)$$

where $\Delta m^2$ is the mass squared difference of the neutrino's proper mass states, and $\theta$ is the mixing angle of the neutrino states, which determines the amplitude of the oscillations. Through oscillations, electron neutrinos pass into sterile states.

The aim of the new experiment is to confirm the oscillation hypothesis and determine the parameters ($\Delta m^2$, $\sin^2 2\theta$) of these oscillations.

If the hypothesis of short-base oscillations is correct, then under conditions where the real parameters of the oscillations are in the sensitivity region of the new gallium experiment (see Section 12), these parameters will be determined with an error of several tens of percent at a significance level of 3σ. Including the parameter $\Delta m^2$, the determination of which is not available in most experiments.

In the new experiment, the sensitivity region for determining the oscillation parameters is continuous in the range $\Delta m^2$ from 0.5 to 5.5 eV$^2$. Therefore, if the obtained neutrino capture rates from the source in different zones do not statistically differ, then this will mean either the oscillation hypothesis is incorrect, or a large value of the mass squared difference ($\Delta m^2 > 5.5$ eV$^2$).

An interesting feature of the experiment is that with three or more independent target zones, BEST-2 will be the first experiment in which an oscillatory curve of the counting rate versus distance can be observed for neutrinos with a fixed energy. Until now, all experiments on neutrino oscillations have observed only a change in the counting rates in the detectors relative to the expected ones, but no experiment could observe any periodicity. In our case, a periodic change in the number of captures versus distance of the type "few-many-few" or "many-few-many" can be observed. Here "few" and "many" mean the relative number of captures sequentially in three target zones.

Another goal of the experiment is to study the dependence of the gallium anomaly on the neutrino energy. In the sterile oscillation hypothesis, such dependence should not exist. This goal is facilitated by the use of a source with a neutrino energy twice as high as in previous gallium experiments. In previous gallium experiments, neutrinos from the sources had approximately the same energy (750 keV in $^{51}$Cr sources and 814 keV in $^{37}$Ar source ). The results of all gallium experiments with sources agree at a level of better than 1.4σ [25]. An indication of the existence of a dependence of the gallium anomaly on neutrino energy may be, for example, a difference in the obtained neutrino capture rate in the new experiment by > 2σ from the average capture rate obtained in previous gallium experiments with sources. In the works [29,30] an experiment was proposed with a $^{65}$Zn source, the neutrino energy in which was 1.8 times higher than in the chromium source. However, the production and use of a zinc source, according to our estimates, is more labor-intensive and expensive than in the case of a $^{58}$Co source. Given that the sensitivity of the experiment to determining the oscillation parameters, as well as to changes in neutrino energy, with a cobalt source is somewhat higher.

### 4. Features of the new experiment

The new experiment is a natural continuation of the BEST experiment [25], which confirmed the gallium anomaly. All procedures of the BEST -2 experiment repeat the procedures of the BEST experiment. A compact neutrino source will be placed in the center of a gallium target divided into zones, in which the neutrino capture rates will be measured independently. Several neutrino irradiations of the target from the source of equal duration will be carried out, and after each one the resulting $^{71}$Ge atoms will be extracted and their decays will be counted in separate counters for each zone of the target.

BEST-2 experiment differs from the previous one in that it will include:

1) due to the introduction of the third target zone, the sensitivity to the distances *L* between the points of emission and capture neutrino has been increased,

and also

2) a neutrino source with higher energy than in the BEST experiment was used .

This difference will make it possible to determine the oscillation parameter $\Delta m^2$ in a wide range of values. The second oscillation parameter, $\sin^2 2\theta$, determines the suppression of the capture rate in the target as a whole, and is currently known with satisfactory accuracy [25] ($\sin^2 2\theta$ ~ 0.4±0.2 at 95% CL). In the new experiment, one can expect an improvement in the precision of its determination by up to 1.4 times (doubling the statistics will improve the statistical error by ~ $\sqrt{2}$ times).

A unique feature of the BEST-2 experiment is the ability to study the dependence of the gallium anomaly on the neutrino energy. If the capture rate measured in the BEST-2 experiment differs from that obtained in previous gallium source experiments, this will mean that there is a dependence of the gallium anomaly on the neutrino energy, which means the need for new physics, but sterile oscillations are not the main cause of the gallium anomaly.

### 5. Source for a new experiment

BEST-2 experiment, which was originally proposed with a $^{65}$Zn source [29,30], will use a $^{58}$Co monochromatic neutrino source.

It is assumed that the $^{58}$Co source will be produced by irradiating nickel in a fast neutron reactor using the reaction $^{58}_{28}Ni(n,p)^{58}_{27}Co$, i.e. the main mass of the active part of the source will consist of nickel (see Sec. 10).

The decay diagram of the $^{58}$Co isotope is shown in Fig. 2.

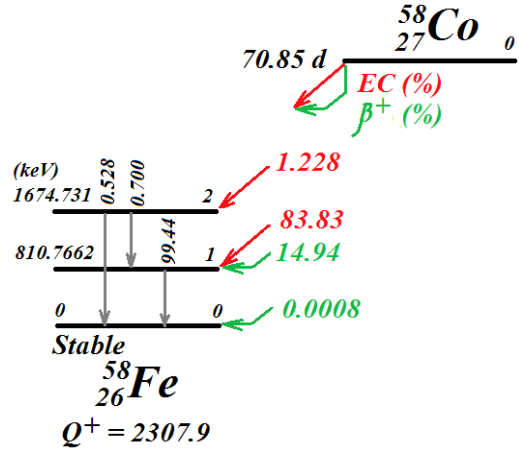

Fig.2. Decay scheme of $^{58}$Co.

$^{58}$Co source emits monochromatic neutrinos with an energy of 1497 keV (98.8%) with a small addition of 633 keV neutrinos (1.2%). The capture cross section of such neutrinos on $^{71}$Ga is σ = 253×10$^{-46}$ cm$^2$, 4.4 times larger than for neutrinos from $^{51}$Cr [31]. The error in the cross section (1.0 $^{+0.17}_{-0.07}$) is taken from J. Bahcall estimates for the energies of the *pep* line neutrinos from the Sun (1442 keV) [32].

Recall that the $^{65}$Zn source emits neutrinos with energies of 1350 (47.8%) and 235 (50.7%) keV [29,30]. Due to the lower energy and lower neutrino yield, the activity of such a source should be significantly higher for comparable statistics.

## 6. Dividing the Ga target into zones

In the new experiment, to determine the oscillation parameter $\Delta m^2$, it is necessary to measure the capture rates at different neutrino path lengths, so it is important to increase the spatial resolution of the detector. This can be achieved by increasing the number of gallium target zones, which will be at different distances from the source. The problem with dividing the target into zones is that it greatly increases the cost of the experiment. Accordingly, the number of zones increases the number of counters, counting channels and systems for pumping gallium from the target zones to chemical reactors for extracting the resulting $^{71}$Ge. At the same time, since the mass of gallium in the target is fixed, then due to the decrease in the average path in the zone of smaller thickness, the statistics in specific zones decreases. Because of this, it is necessary to use a more intense source. The simplest option for increasing the number of target zones is to divide the existing two-zone target, which was used in the BEST, into two parts with a cylindrical shell with an approximately equal thickness of gallium. The average thickness of gallium in the three zones of the target will be approximately 52, 27 and 27 cm. The statistics of the first zone in this case will be twice as large as in the second and third zones. The same statistics can be obtained by dividing the inner, spherical zone into two equal parts in thickness. However, in this case the mass of gallium in the zones is too small for the operation of the system for extracting $^{71}$Ge atoms from gallium.

Note that if we divide the gallium target into 4 zones with equal gallium thickness, the statistics in the two outer zones will not change, and in the two inner zones it will be equal to the statistics in the two outer zones. Since the amount of gallium in the two inner zones is less than in the two outer zones, the background from solar neutrinos in them will be less. Therefore, making 4 zones seems more advantageous than making 3 zones. However, as our analysis showed, using four zones in the experiment with a fixed total target mass and source activity does not lead to a noticeable change in the sensitivity of the experiment to determining the oscillation parameters. Therefore, it is proposed to use 3 zones of the gallium target in the BEST-2 experiment (Fig. 3).

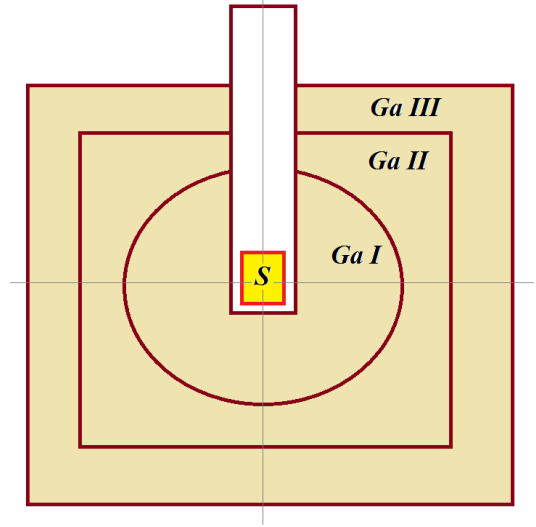

Fig. 3. Schematic diagram of the gallium target in the BEST-2 experiment. The $^{58}$Co(S) source is placed in the common center of all three target zones through a vertical pipe. The outer boundaries of the second and third target zones (Ga II and Ga III) are made in the form of vertically arranged cylinders

The approximate dimensions of the target zones will therefore be as follows: the inner radius of the spherical target is 67 cm; the inner radius and height of the middle cylindrical shell of the target is 83 cm and $83 \times 2 = 166$ cm; the inner radius and height of the outer cylinder limiting the third zone is 109 cm and 234 cm.

Figure 4 shows the probability distributions of neutrino capture in gallium as a function of the distance between the emission points in the source and the capture in three target zones, as well as the total distribution for all zones. The distributions were obtained by the Monte Carlo method for a source with a uniformly emitting part in the form of a cylinder with a diameter of 14 cm and a height of 26 cm (volume of 4 l), located in the common center of the target zones. Such a volume can accommodate up to 36 kg of nickel, from which the cobalt source is made. The amount of nickel required to manufacture the source will most likely be less, so the estimates of the sensitivity of the experiment given below are conservative.

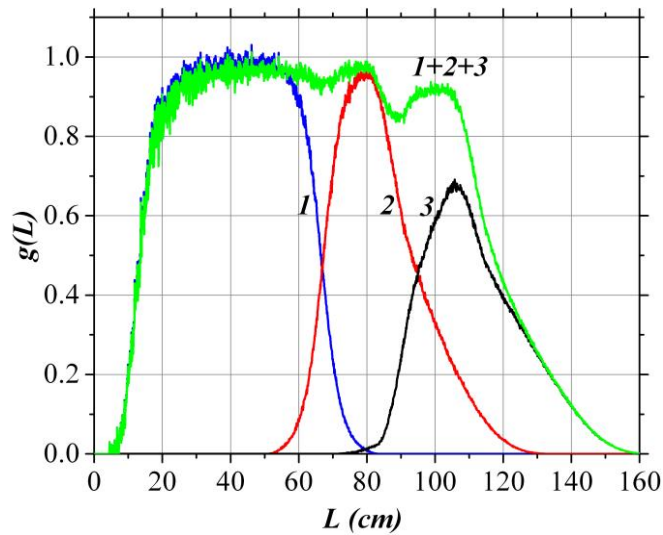

Fig. 4. Distributions of neutrino capture probabilities by gallium depending on the distance $L$ between emission and capture points in three target zones in relative units.

## 7. Effect of oscillation parameters on counting rates

In this experiment, the hypothesis of electron neutrino oscillations into sterile states with large values of the parameter $\Delta m^2$ (~1 eV$^2$) will be investigated. Note that here transitions into any states can be investigated, including antineutrinos, since only electron neutrinos are registered on the gallium target.

Oscillations reduce the neutrino capture rates in the gallium target. The decrease in the capture rates depends on the oscillation parameters $\Delta m^2$ and $\sin^2 2\theta$ in accordance with the survival probability (1). For monochromatic neutrinos, the value of $P_{ee}$ has a sinusoidal dependence on the distance $L$ between the points of neutrino production and absorption. In the gallium target, it is not possible to single out each event by distances, and the number of neutrino captures is summed up over all distances within each zone. Therefore, the survival probabilities are also averaged over all distances within the zones.

For three zones, the dependences of the expected capture rates in different zones of the target on the parameter $\Delta m^2$ are shown in Fig. 5 for the value $\sin^2 2\theta = 0.30$.

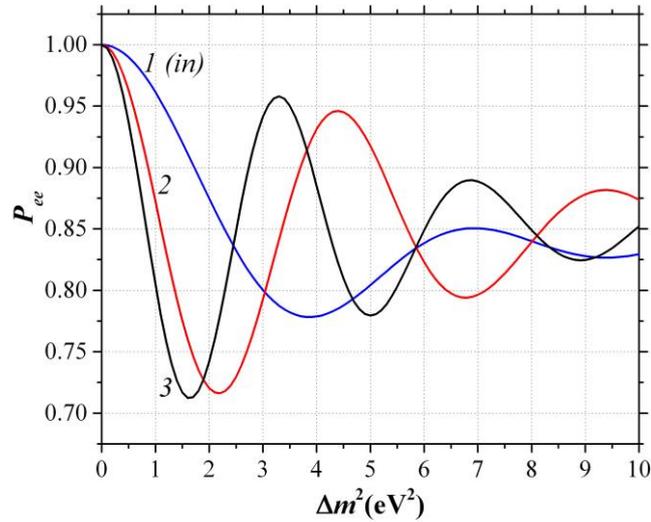

Fig. 5. Neutrino capture rates in three zones of the BEST-2 target depending on the parameter $\Delta m^2$ for a fixed value of the parameter $\sin^2 2\theta = 0.30$

The capture rates in different target zones change differently with the change of the parameter $\Delta m^2$. The ratio of the neutrino capture rates in different target zones is shown in Fig. 6.

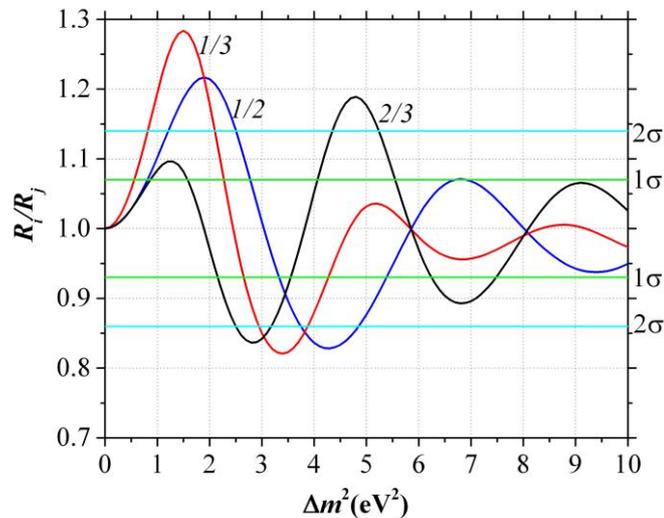

Fig. 6. Pairwise ratios of neutrino capture rates in three zones of the BEST-2 target depending on the parameter $\Delta m^2$ for a fixed value of the parameter $\sin^2 2\theta = 0.30$

It is evident that in the range of $\Delta m^2$ from ~0.5 to 6 eV$^2$ the ratios of the capture rates differ significantly, which allows us to determine the parameter $\Delta m^2$ if it is in the specified range of values. For a given value of the oscillation amplitude $\sin^2 2\theta$, the figure also shows the limitations on the ratios of the capture rates in different target zones ±1σ and ±2σ, according to which the sensitivity region of the experiment to the determination of the oscillation parameters is built (see Section 12).

Determination of the parameter $\Delta m^2$ is possible in the regions of $\Delta m^2$, in which the difference in the counting rates in different zones of the target will be larger than the measurement errors of these rates. The ratio of the counting rates in different zones becomes close to unity for values of $\Delta m^2 > 10$ eV$^2$. Figure 6 also shows blind zones, in which the ratios of all count rates are equal to unity – near the values of $\Delta m^2 = 6$ and 8 eV$^2$. In such regions, determination of the parameter $\Delta m^2$ will be impossible in this experiment.

**8. Exposure time**

BEST-2 experiment will collect events according to the scheme of other gallium experiments. The measurement cycle consists of several procedures: 1) irradiation of the gallium target with neutrinos from the source; 2) extraction of $^{71}$Ge atoms produced in neutrino interactions; 3) transfer of the extracted germanium into gas (germane - GeH$_4$) and placing it in a gas proportional counter; 4) counting of $^{71}$Ge decays in a proportional counter. To increase the statistics of events, several measurement cycles $m$ are carried out. Usually $m = 10$. The exposure and extraction times remain the same in all measurements. In addition to increasing the statistics, repeated measurements lead to a decrease in some systematic errors [25].

Figure 7 shows the number of events in the target spherical zone of the target when irradiated with a $^{58}$Co source with an activity of 0.40 MCi, depending on the duration of one exposure $t_1$, which is the same for all irradiations.

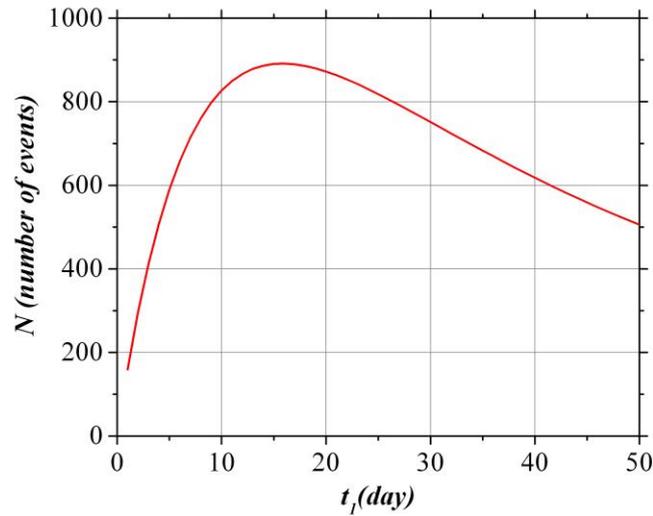

Fig. 7. The number of events from the decay of $^{58}$Co depending on the duration of one exposure $t_1$ for an initial activity of 400 kCi.

The figure shows the dependence of the number of events on the time of one exposure $t_1$:

$$N(t_1) = \frac{p}{\lambda_1 - \lambda_0} \cdot (e^{-\lambda_0 t_1} - e^{-\lambda_1 t_1}) \cdot \frac{1 - e^{-\lambda_1(t_1+t_2)\cdot m}}{1 - e^{-\lambda_1(t_1+t_2)}} \cdot \varepsilon \qquad (2)$$

Here $\lambda_0 = \dfrac{\ln 2}{11.43d}$ and $\lambda_1 = \dfrac{\ln 2}{70.86d}$ are the decay constants of $^{71}$Ge and the source isotope $^{58}$Co; $p$ is the neutrino capture rate from the source at the beginning of the first exposure; $m$ is the number of irradiations; $\varepsilon = 0.5$ is the total efficiency of all experimental procedures or the ratio of the number of registered decays and the number of $^{71}$Ge atoms in the target at the time of extraction. For estimates, we fix the number of exposure $m = 10$. Time $t_2 = 1$ day is the interval in exposures for extractions and activity measurements.

The maximum statistics in the experiment is achieved at the extreme point, with the value $t_1 = 16$ days. For a source activity of 400 kCi, the total number of registered pulses is expected to be $N = 891$.

### 9. Source activity

To estimate the sensitivity of the experiment with the $^{58}$Co source, the statistics of the BEST experiment [25] were used. In the BEST experiment, the statistics of events in one zone was approximately 700 events. Therefore, in the new experiment, the activity of the source should be such that the number of events in the first target zone was approximately 700, and in the other two – 350 each.

The rate of neutrino capture from a monochromatic source is calculated using the formula [8]:
$$p = A \cdot D \cdot <L> \cdot \sigma, \qquad (3)$$
where $A$ is the source activity; $D = 2.1 \times 10^{22}$ at $^{71}$Ga/cm$^3$ is the density of $^{71}$Ga atoms in the target; $<L> = 53$ cm is the average neutrino path length in Ga in one zone of the BEST target; $\sigma = 253 \times 10^{-46}$ cm$^2$ is the neutrino capture cross section by $^{71}$Ga nuclei for the $^{58}$Co source.

The amount of extracted $^{71}$Ge is determined by formula (2) for the values of the time of one irradiation $t_1 = 16$ days and the time interval between adjacent exposures $t_2 = 1$ day.

Considering that oscillations suppress the counting rate by 20%, the number of events in the first target zone should be $N \sim 700/0.8 = 875$.

Figure 8 shows the dependence of the amount of registered atoms of $^{71}$Ge $N$ from the number of exposures $m$ (formula (2)). With the initial neutrino capture rate $p = 36$ days$^{-1}$, it is possible in $m = 10$ irradiations to achieve the required number of events in the first (spherical) zone, $N = 891$.

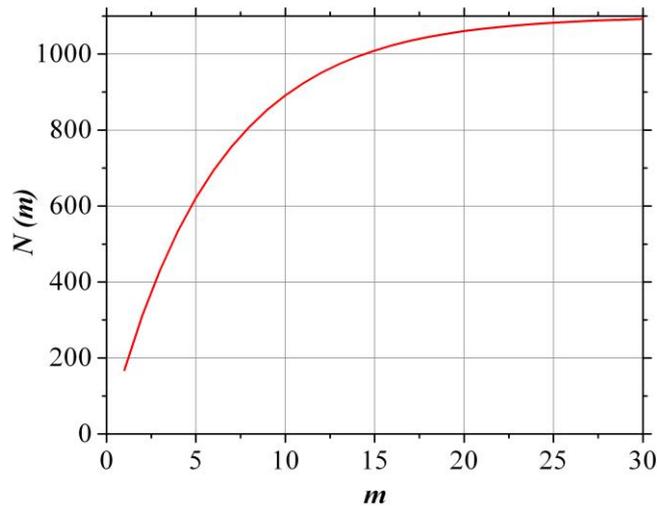

Fig. 8. Dependence of the amount events $N$ on the number of exposures $m$ for the experiment with the $^{58}$Co source.

To obtain such statistics, the activity of the $^{58}$Co source is required $A = 400$ kCi.

Note that with the source activity almost an order of magnitude less than in the BEST experiment (3.4 MCi), the number of events in the new experiment will be the same as in BEST. The reason for this is as follows:

1) at higher neutrino energy, the capture cross section in the new experiment is 4.4 times higher;

2) a longer lifetime of the source in the new experiment ($T_{1/2}$ = 71 days versus 27.7 days in the BEST), due to which later irradiations make a higher contribution. Therefore, the duration of exposures increases ($t_1$ = 16 days versus 9 days in the BEST).

## 10. Source production

A source of $^{58}$Co can be produced in a fast neutron reactor by the reaction

$$^{58}_{28}Ni(n,p)^{58}_{27}Co \qquad (4)$$

For neutrons with an energy of 14-15 MeV, the cross section of reaction (4) can be estimated using the scheme proposed in [33], and it is equal to σ = 0.34 b. Taking into account the wide spectrum of neutrons in fast neutron reactors, the cross section of such a reaction will probably be less: σ = 0.1439 b [34]. In further estimates, this value of the cross section will be used.

The formation of the isotope $^{58}$Co in a constant flux of fast neutrons occurs according to the equation:

$$N_{Co}(t) = \frac{N_{Ni}(0) \cdot \Phi \cdot \sigma}{\Phi \cdot \sigma - \lambda} \cdot (e^{-\lambda t} - e^{-\Phi \sigma t}) \qquad (5)$$

Here $N_{Ni}(0)$ is the initial number of $^{58}$Ni atoms in the neutron flux $\Phi$; σ is the cross section of neutron capture by the $^{58}$Ni isotope; λ is the decay constant of $^{58}$Co. Accordingly, the cobalt activity will be $A(t) = \lambda \cdot N_{Co}(t)$. The fast neutron flux in the BOR60 reactor (RIAR) reaches $3.7 \cdot 10^{15}$ cm$^{-2}$ s$^{-1}$ [35], and in the BN600 reactor – $2.3 \cdot 10^{15}$ cm$^{-2}$ s$^{-1}$.

For estimates, we will consider reach the activity of $^{58}$Co in a fast neutron flux $\Phi = 2 \cdot 10^{15}$ cm$^2$s$^{-1}$. Figure 9 shows curve of reach the activity for 15 kg of natural nickel under conditions where the resulting $^{58}$Co does not burn out, i.e. does not interact with neutrons in the reactor.

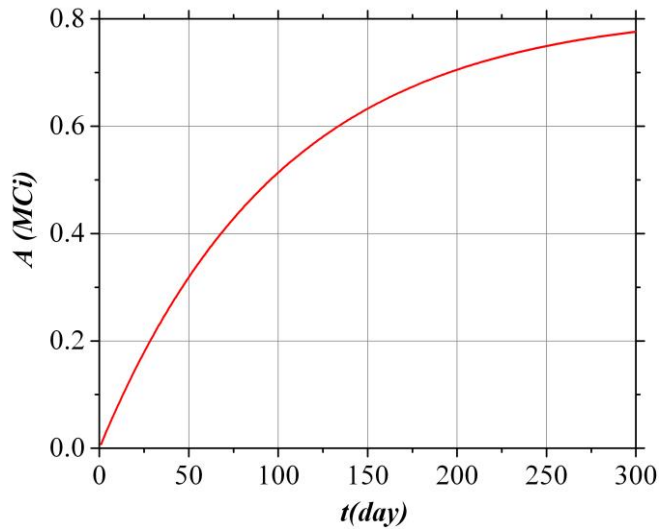

Fig. 9. Curve of reach the $^{58}$Co activity in a fast neutron flux $\Phi = 2 \cdot 10^{15}$ cm$^{-2}$s$^{-1}$ from 15 kg of natural nickel.

$^{58}$Co 0.4 MCi required for the experiment is accumulated in approximately 70 days of irradiation.

The isotope $^{58}$Ni content in natural nickel is 68.27%, and in 15 kg of natural nickel there will be $N_{Ni}(0) = 1.1 \cdot 10^{26}$ atoms of $^{58}$Ni. Neutron captures by other nickel isotopes do not lead to the producing of radioactive isotopes, except for the reaction $^{60}_{27}Ni(n,p)^{60}_{28}Co$. The cross section of this reaction is $\sigma_{60} \approx 0.040$ b [33]. Taking into account the natural content of $^{60}$Ni (26.1%), the rate of producing of $^{60}$Co ($v_{prod} \sim f \cdot \sigma$) is less than the rate of the main isotope ($^{58}$Co) production by about 9.4 times. Accordingly, the decay rates of isotopes in the source ($v_{decay} \sim \lambda \cdot v_{prod} \cdot t$) differ by ~250 times.

For nickel enriched in isotope 68, the same activity can be expected to be produced in a reactor with a nickel mass 30% smaller, i.e. about 10 kg. The volume of the emitting part in this case can be about 1.2 l, only 2 times larger than the volume of the emitting part of the chromium source in the BEST experiment.

The enrichment of nickel for the source can also lead to additional purification of the material from elements, producing radioactive impurities, which increase the error in measuring the activity of the source by the calorimetric method. For example, in the BEST experiment, during the enrichment of chromium, a material purity was achieved at which the error in measuring the thermal power of the source from radioactive impurities was negligibly small (~ $5 \cdot 10^{-6}$) [36].

## 11. Statistical errors of measurements

Let us estimate the statistical errors of gallium measurements with a source. We will assume that the number of extracted $^{71}$Ge atoms has a Poisson distribution, in which the statistical error is equal to the square root of the number of events $\sigma = \sqrt{N}$. The number of counted $^{71}$Ge decays also has a Poisson distribution. But due to the counter background, the error will be greater: $\sigma = \alpha \cdot \sqrt{N}$, where $\alpha > 1$.

Let us consider the error for one target zone. After $m$ exposures we obtain $m$ sets $\{N_i\}$ of registered numbers of $^{71}$Ge decays in proportional counters. The total number of events and its error are equal to $N = \sum_{i=1}^{m} N_i$ and $\sigma = \alpha \cdot \sqrt{\sum_{i=1}^{m} N_i}$. Decaying $^{71}$Ge is produced from the source and from solar neutrinos, which are the only background in the experiment. We will assume that the same number of $^{71}$Ge – $N_{Sun}$ atoms is formed from the Sun in each exposure of duration $t_1$. Then the total number of events from the source is equal to $N_S = \sum_{i=1}^{m} N_i - m \cdot N_{Sun}$, and the error in this number is equal to $\sigma = \alpha \cdot \sqrt{\sum_{i=1}^{m} N_i + m \cdot N_{Sun}}$. The relative error will be $\delta = \frac{\sigma}{N_S} = \alpha \cdot \frac{\sqrt{N_S + 2 \cdot m \cdot N_{Sun}}}{N_S}$.

For $\alpha = 1$ we estimate the errors for a 3-zone target. In the 3-zone target in the inner spherical zone the number of events in the new experiment is equal to the number of events in the same zone in the BEST, i.e. $N_S = 700$, and in the other two it is 2 times less, i.e. 350 each. Here we have already taken into account that the counting efficiency of the extracted $^{71}$Ge atoms is ε=0.5 [25]. The number of registered events from the Sun is $N_{Sun} = \frac{v_{Sun}}{\lambda} \cdot (1 - e^{-\lambda t}) \cdot \varepsilon$ [12]. For a Ga target with a mass of 50 t, the neutrino capture rate from the Sun is $v_{Sun} = 1$ day$^{-1}$ (66.1±3.1 SNU) [12]. λ is the decay constant of $^{71}$Ge produced in the target ; $t$ is the duration of target exposure with solar neutrinos until the next extraction.

Table 1 shows the masses of the target zones M (in tons), the number of expected events from the source $N_S$ and from solar neutrinos $N_{Sun}$ for the duration of one exposure $t = t_1$, as well as the relative statistical errors δ.

Table 1

| Zone No. | M, t | $N_S$ | $N_{Sun}$ | δ (α=1) |
|---|---|---|---|---|
| 1 | 7.7 | 700 | 0.9 | 0.038 |
| 2 | 14.7 | 350 | 1.7 | 0.056 |
| 3 | 26.8 | 350 | 3.1 | 0.058 |

The statistical errors in zones 2 and 3 are slightly different due to the different masses of gallium in the target zones, which causes the contribution of solar neutrinos in these zones to differ.

Based on the experience of measuring solar neutrinos, for the number of registered $^{71}$Ge decays greater than 10 in the L or K peaks, the value of α does not exceed 1.1. Note that events from decays of $^{71}$Ge in proportional counters form two peaks in the pulse spectrum (L and K peaks) in the energy range of 1.2 and 10.4 keV [9]. For a $^{58}$Co source with an activity of 400 kCi, in all exposures, including the last, tenth, the number of events in the L and K peaks of $^{71}$Ge decays will be obviously larger: in the 10th exposure, the expected number of pulses in each peak will be equal to

$$N_{L(K)} = \frac{p}{\lambda_0 - \lambda_1} \cdot (e^{-\lambda_1 \cdot t_1} - e^{-\lambda_0 \cdot t_1}) \cdot e^{-\lambda_0 \cdot (t_1+t_2) \cdot (m-1)} \cdot \varepsilon \cdot P_{GA} \cdot P_{2,3} = 8.$$

Here $p = 36$ days$^{-1}$ is the neutrino capture rate in the inner zone of the target at the beginning of the first exposure; $\lambda_0$ and $\lambda_1$ are the decay constants of $^{58}$Co and $^{71}$Ge; $m = 10$ is the number of exposures; $t_1$ and $t_2$ are the time of one exposure and the interval between exposures; $\varepsilon = 0.5$ is the efficiency of recording the pulse from the decay of the $^{71}$Ge atoms extracted from the target ; $P_{GA} = 0.8$ is the expected suppression of the number of $^{71}$Ge atoms due to the gallium anomaly; $P_{2,3} = 0.5$ is the decrease in the counting rate in the outer zones of the target due to their smaller thickness compared to the thickness of the inner (spherical) zone.

Therefore, one can expect that for the total number of events the value $α ≈ 1$.

Table 2 presents the expected statistics, the number of registered (i.e. the total number of events in the *L* and *K* peaks) events ($n_i$), in the large, spherical, target zone in each exposure of the gallium target by a $^{58}$Co source with an activity of 400 kCi. The expected numbers of events in the other two target zones are equal to half of the given ones, respectively, to the average neutrino ranges in Ga in these zones. The exposures are carried out with the values $t_1 = 16$ days and $t_2 = 1$ day; solar neutrino captures are not considered. Information on additional exposures, in excess of the adopted number $m = 10$, may also be useful for the estimates.

The first line of Table 2 contains the exposure number. The second line contains the values of the end times of the exposures relative to the beginning of the first exposure. The last line contains the summary statistics for $m = 10$ and 14 exposures.

Table 2

| No. | 1 | 2 | 3 | 4 | 5 | 6 | 7 | 8 | 9 | 10 | 11 | 12 | 13 | 14 |
|---|---|---|---|---|---|---|---|---|---|---|---|---|---|---|
| T,d | 16 | 33 | 50 | 67 | 84 | 101 | 118 | 135 | 152 | 169 | 186 | 203 | 220 | 237 |
| $n_i$ | 169 | 143 | 121 | 102 | 87 | 73 | 62 | 53 | 45 | 38 | 32 | 27 | 23 | 19 |
| Σ($n_i$) | | | | | | | | | | 891 | | | | 993 |

**12. Sensitivity to the oscillation parameters determination**

Figure 10 shows the sensitivity regions of the BEST-2 experiment with a $^{58}$Co source with an activity of 400 kCi for a 3-zone gallium target for determining the parameter $\Delta m^2$.

The regions of sensitivity to the determination of oscillation parameters were determined by the ratios between the capture rates in different zones of the gallium target at various values of the oscillation parameters.

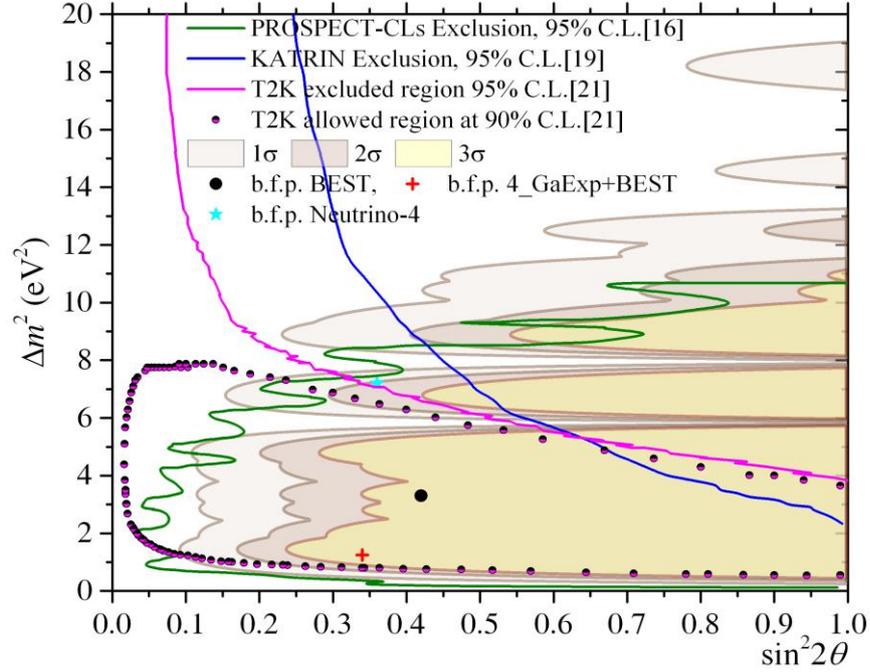

Fig.10. Sensitivity regions to determining the parameter $\Delta m^2$ on a 3-zone target with an inner sphere. The figure also shows the parameter regions excluded by the analysis of the PROSPECT [16], KATRIN [19] and T2K [21] experiments data (to the right of the constraint curves), and also shows the region of allowed parameters obtained in the T2K [21] experiment.

In the absence of oscillations, the expected neutrino capture rate in the target zone $i$ is
$$v_{0i} = A \cdot \sum_n g_i(L_n)$$

Here $A$ is a factor including the source activity, the neutrino capture cross section in gallium, and the density of gallium atoms in the target. The summation is performed over all lengths $L$ by the values of the function $g_i(L)$ – the probability distributions of neutrino capture in gallium in the target zone $i$ over the distances shown in Fig. 4.

Under oscillation conditions, the expected capture rates will be $v_{ilk} = A \cdot \sum_n g_i(L_n) \cdot P_{ee\_n\_lk}$

Here $P_{ee\_n\_lk} = P_{ee\_n}(\Delta m^2_l, \sin^2 2\theta_k)$ is the probability of survival of electron neutrinos at a distance $L_n$ from the birth point under conditions of oscillations with parameters $(\Delta m^2, \sin^2 2\theta) = (\Delta m^2_l, \sin^2 2\theta_k)$[1].

---

[1]For a source with two neutrino lines, such as the $^{58}$Co source, the survival probability is calculated using the formula

$P_{ee} = P_1 \cdot \dfrac{1 + \alpha \dfrac{P_2}{P_1}}{1 + \alpha}$, where $\alpha = \dfrac{f_2 \cdot \sigma_2}{f_1 \cdot \sigma_1}$; $P_1$ and $P_2$ are the survival probabilities of neutrinos with energies $E_1$ and $E_2$, and $f_i$ and $\sigma_i$ are the yields of neutrinos of such energies in decays of the source isotope and their capture cross sections in gallium. For $^{58}$Co $E_1 = 1497$ keV, $f_1 = 0.988$, $\sigma_1 = 253 \times 10^{-46}$ cm$^2$ and $E_2 = 633$ keV, $f_2 = 0.012$, $\sigma_2 = 46.5 \times 10^{-46}$ cm$^2$, and $\alpha = 0.00255$.

In the analysis we used the relative neutrino capture rates in the $i$-zone of the target under oscillation conditions: $R_{ilk} = \dfrac{v_{ilk}}{v_{0i}}$.

For each point of the oscillation parameters $(l,k)$ we find the differences $D(l,k) = \max\limits_{\substack{i,j=1,2,3; \\ i<j}} |1 - \dfrac{R_{ilk}}{R_{jlk}}|$.

Thus, the value of $D(l,k)$ is determined by the maximum difference between the expected capture rates in different target zones for the given values of the oscillation parameters. An example of capture rates for different values of the parameter $\Delta m^2$ at a fixed value of $\sin^2 2\theta$ is shown in Fig. 5.

The boundaries of the sensitivity regions were built by the elements of the two-dimensional matrix $D(l,k)$ (which was calculated for the values of $\Delta m^2$ with a step of 0.05 eV$^2$ and for $\sin^2 2\theta$ with a step of 0.002) for $D(l,k) = 1\sigma$, $2\sigma$ and $3\sigma$, where the standard deviation $\sigma = 7\%$. The value of 7% approximately corresponds to the statistical error of the expected result of measuring the counting rate in one of the outer zones (2nd or 3rd) in the absence of oscillations. Thus, to build the sensitivity regions, we used a simplified scheme with a fixed statistical error. Such simplification has virtually no effect on determining the boundaries of the sensitivity regions, since the error changes noticeably only at large oscillation amplitudes, i.e., in the region of large values of $\sin^2 2\theta$.

The curves built in this way restrict the regions within which the values of the oscillation parameters ($\Delta m^2$, $\sin^2 2\theta$) can be definitely determined for significance levels of $2\sigma$ and higher. In this case, the regions of allowed oscillation parameters form a single compact region with uncertainties of the order of several tens of percent. Outside the sensitivity zone, the regions of allowed values of the parameter $\Delta m^2$ are divided into many separated regions with different values of $\Delta m^2$, and for small amplitudes of $\sin^2 2\theta$, the resulting regions of allowed values of $\Delta m^2$ are distributed continuously from some small value to infinity. In such cases the parameter $\Delta m^2$ cannot be determined. Examples of regions of allowed oscillation parameters that can be obtained from the experimental results are shown in Figs. 11 and 12.

Figure 10 also shows a curve excluding the region of oscillation parameters, according to the PROSPECT experiment data [16]. It is evident that these data almost completely exclude the region of oscillation parameters to which the new gallium experiment is sensitive. Part of the region of sensitivity to determining the allowed parameters of the BEST-2 experiment is also excluded by the data of the T2K experiment [21]. At the same time, the region of allowed oscillation parameters obtained in the same T2K experiment [21] can be tested in the BEST-2 experiment. In addition, in [23], the regions of allowed parameters of oscillations into sterile states of muon neutrinos observed in the IceCube experiment were obtained, including the parameter values 2.4 eV$^2 < \Delta m^2 < 9.6$ eV$^2$ at a significance level of 90%. In Fig. 10, the IceCube limitations are not shown, since the oscillation amplitudes in [23] are given in other units ($|U_{\mu 4}|^2$ and $|U_{\tau 4}|^2$).

Fig. 11 shows an example when it is assumed that the oscillation parameters are within the sensitivity region. In the figure, the region of allowed parameters is localized in a limited range of parameters.

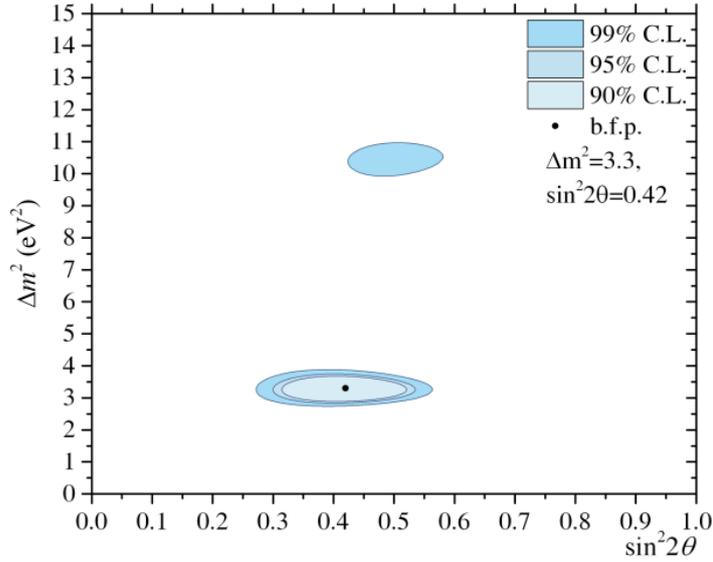

Fig.11. An example of the regions of allowed oscillation parameters that are within the sensitivity region of the experiment: ($\Delta m^2$, $\sin^2 2\theta$) = (3.3 eV$^2$, 0.42) at the measured count rates ($R_1$, $R_2$, $R_3$) = (0.70, 0.78, 0.94)

An example of when the oscillation parameters are outside the sensitivity region to determining of the oscillation parameters is shown in Fig. 12. Here the parameter $\Delta m^2$ can take any value from ~6 eV$^2$ to infinity with almost equal probability.

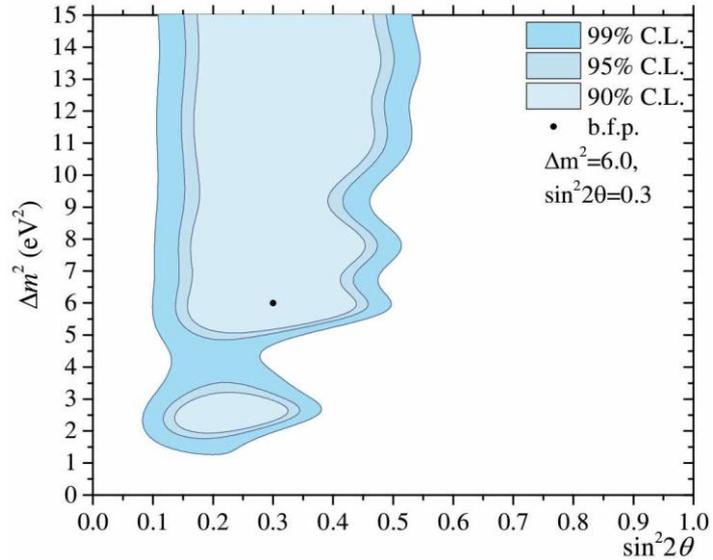

Fig. 12. An example of the regions of allowed oscillation parameters that are not within the sensitivity region of the experiment: ($\Delta m^2$, $\sin^2 2\theta$) = (6.0 eV$^2$, 0.3) at the measured count rates ($R_1$, $R_2$, $R_3$) = (0.84, 0.82, 0.85)

### 13. Sensitivity regions for determining oscillation parameters

The regions of allowed values of the oscillation parameters are determined by the function $\chi^2(\Delta m^2, \sin^2 2\theta) = (\vec{R}_{meas} - \vec{R}_{calc})^T V^{-1} (\vec{R}_{meas} - \vec{R}_{calc})$ [24], where $\vec{R}_{meas}$ and $\vec{R}_{calc}$ are the vectors of the ratios of the measured and calculated counting rates of the studied data sets with oscillation parameters ($\Delta m^2$, $\sin^2 2\theta$) to the expected counting rates in the absence of oscillations; $V$ is the covariance matrix, including statistical and systematic, including uncorrelated, experimental errors. In the region of the studied parameters, we find the most probable (BF, best fit ) the value of ($\Delta m^2$, $\sin^2 2\theta)_{bf}$, at which the function $\chi^2 = \chi^2_{min}$ takes a minimum value. The regions of allowed parameters are shown in Fig. 11 and 12 and are determined by the values of the parameters, which are found from the inequality $\chi^2 < \chi^2_{min} + \Delta\chi^2$ for distributions with two degrees of freedom, where $\Delta\chi^2 = 2.30, 6.18, 11.83$ for significance levels of $1\sigma$ (68.27%), $2\sigma$ (95.45%), $3\sigma$ (99.73%).

Figure 10 shows that both BF values obtained in the BEST experiment – from the BEST experiment data and from the data of all gallium source experiments – are within the range of parameter values where the parameters can be determined with a significance level of $3\sigma$: $(\Delta m^2, \sin^2 2\theta) = (3.3, 0.42)$ and $(1.25, 0.34)$.

BF values of the oscillation parameters obtained in the Neutrino-4 experiment, $(\Delta m^2, \sin^2 2\theta) = (7.2, 0.36)$, are at the boundary of the sensitivity region of the BEST-2 experiment with the $^{58}$Co source (at the region boundary of the $2\sigma$ significance level). In the experiment with the $^{65}$Zn source with a lower energy of emitted neutrinos (1350 keV instead of 1497 keV), this region is insensitive to the central value of the Neutrino-4 result.

## 14. Potential results of the BEST-2 experiment

Let us consider what physical results can be obtained in the BEST-2 experiment. To do this, we will set two parameters – 1) the total counting rate for all target zones $R_0$ and 2) the maximum difference between the counting rates in different target zones $\Delta R$. The total counting rate $R_0$ will be compared with the counting rate $R = 0.80\pm0.05$, obtained from the results of all previous gallium experiments with sources.

Depending on these two parameters, the result of the experiment can be classified into one of three options:

1) If $R_0 \approx R$ and $\Delta R > 2\sigma$, i.e. the obtained counting rate approximately coincides with the rate of previous gallium experiments with sources and a significant difference in the counting rates in different zones of the target is observed, this will mean that the gallium anomaly is associated with oscillations into sterile states and the parameters of these oscillations will be determined in this experiment, i.e. the oscillation parameters are within the sensitivity region for determining the parameters.

2) If $R_0 \approx R$ and $\Delta R < 2\sigma$, i.e. the difference in counting rates in different target zones is small, then sterile oscillations remain a possible solution to the gallium anomaly problem, the gallium anomaly will be confirmed at a higher statistical significance level, but the oscillation parameters will not be determined, i.e. the oscillation parameters are outside the sensitivity range for determining the parameters.

3) If $R_0 \neq R$, i.e. the measured counting rate will differ significantly from the counting rates in previous gallium experiments, then with any difference in the counting rates between different target zones the dependence of the gallium anomaly on the neutrino energy will be determined. In this case, oscillations to sterile states cannot be the cause of the gallium anomaly, although sterile oscillations can make a minor contribution to the gallium anomaly. And to explain the major contribution to the gallium anomaly, we will have to look for other hypotheses.

## 15. Operation of the source

The neutrino source $^{58}$Co proposed for the BEST-2 experiment will have characteristics that need to be taken into account when working with it. The main ones are heat release and radiation activity. They are significantly higher than the corresponding values of the characteristics of the $^{51}$Cr and $^{37}$Ar sources used in previous gallium experiments.

## 15.1. Heat release of the source

The heat generated at the source will affect the materials of the nuclear reactor and the structural materials of the target zone shielding and shells, which may deform under thermal overloads.

To estimate the heat release of the source, we will assume that all the energy released during the decay of the isotope nuclei is converted into heat, with the exception of the energy carried away by neutrinos. The protective shells of the source absorb all emitted charged particles and all gamma radiation. The estimates neglected the escape of high-energy photons beyond the source's protection, as well as the heat release of impurities, which will most likely contribute less than 1%.

Table 3 shows the photon energies and their yields upon decay of $^{58}$Co [37].

The heat release of the source will be equal to $P = \sum_i E_i \cdot f_i + E_e$. Here $E_e \approx 40$ keV/decay is the energy of Auger electrons. The release of thermal energy will be about 1 MeV/decay and the thermal power of a source with an activity of 400 kCi will be about 2.4 kW.

For comparison, the heat release of the $^{51}$Cr neutrino source with an activity of 3.4 MCi in the BEST experiment was 740 W [25].

The high heat release may require special cooling of the source at time inside the gallium targets, as well as in the setups where its activity will be measured.

Table 3

| $E_\gamma$, keV | 511 | 810.76 | 864 | 1674.7 |
|---|---|---|---|---|
| Output γ line $f$, % | 29.88 | 99.44 | 0.70 | 0.528 |

## 15.2. Radiation activity of the source

Let us evaluate the conditions for safe work with the source from the point of view of radiation exposure of laboratory personnel.

The dose rate of personnel exposure to $^{58}$Co will be estimated by the formula:

dr = A · $\varepsilon_g$ · E$\gamma$ · $D$ · $f$ / $m$

Here A is the activity of the source; $\varepsilon_g$ is the geometric efficiency, i.e. the probability of a photon emerging from the surface of the source's shielding hitting the human body; $E_\gamma$ is the photon energy; $D$ is the factor of radiation transmission through the shielding; $f$ is the radiation yield in the decay of the isotope; $m$ is the mass of an average person.

We assume that in the immediate vicinity of the source, 10 cm from its center, the geometric efficiency is $\varepsilon_g = 0.2$ and the average person's mass is $m = 70$ kg. As the distance from the source increases by a distance $L$, the dose rate decreases proportionally to $1/L^2$.

For the estimates, it was assumed that the diameter of the emitting part of the source would be about 10 cm. Then, since the tube through which the source is lowered into the center of the gallium target (Fig. 3) has a diameter of 22 cm, the source biological shielding can be up to 4 cm thick. The biological shielding – the non-removable shell of the active part of the source – can be made of tungsten,

Table 4 shows the coefficient D values of passing such shielding for γ lines of different elements, calculated according to reference tables from [38].

For a $^{58}$Co source with an activity of 400 kCi, the dose rate directly behind a 4 cm thick tungsten biological shield will be 5.9 mSv/s. This also takes into account the radiation from $^{60}$Co, the activity of which will be about 1% of the initial activity of $^{58}$Co.

Table 4. Photon emission suppression coefficient for several lines of different sources after passing through 4 cm thick tungsten shielding.

| $E_\gamma$, keV | 511 ($^{58}$Co) | 811 ($^{58}$Co) | 864 ($^{58}$Co) | 1115 ($^{65}$Zn) | 1173 ($^{60}$Co) | 1332 ($^{60}$Co) | 1675 ($^{58}$Co) |
|---|---|---|---|---|---|---|---|
| D, 4 cm | 3.2 e-5 | 0.0022 | 0.0029 | 0.0083 | 0.0096 | 0.014 | 0.025 |

With a permissible radiation dose for category A personnel over 1 year $D_0$ = 20 mSv [39], one can be near the source for no more than $t = D_0/dr$ = 3.4 s.

For comparison, one can be near a source of $^{65}$Zn of the same activity (400 kCi) with the same biological shielding for no more than 0.65 s.

Therefore, additional shield is needed to work with the source. Using, for example, 10 cm thick lead bricks as additional protection increases the permissible working time at a distance of 1 m from the source to 10 hours, i.e. it becomes quite realistic for carrying out relatively complex work. With such shield (4 cm W + 10 cm Pb ), the fraction of radiation from $^{60}$Co and from the 1675 keV line in the dose rate increases to ~10%.

Additional lead shielding will be activated only when the source is outside the three-zone target, while moving it between setups where its activity will be measured.

## 16. Conclusion

The new gallium experiment BEST-2 will investigate the gallium anomaly, which may indicate the manifestation of new physics. The division of the BEST-2 gallium target into 3 independent zones and the use of a monochromatic neutrino source $^{58}$Co with an activity of 400 kCi makes the experiment sensitive to the hypothesis of oscillations into sterile states in a wide range of oscillation parameters. When the oscillation parameters are within the sensitivity range of the experiment, the values of both parameters - the amplitude $\sin^2 2\theta$ and the frequency $\Delta m^2$ - will be determined in the experiment. The sensitivity range of the experiment includes the most probable values of the oscillation parameters obtained in previous gallium experiments with artificial neutrino sources, as well as in the Neutrino-4 experiment.

The experiment will involve 10 irradiations of a gallium target over a period of about six months. The experiment's statistics will be comparable to the BEST experiment's statistics , and the errors of both experiments will be comparable.

The energy of neutrinos produced by the $^{58}$Co source is approximately twice as high as that of the $^{51}$Cr and $^{37}$Ar sources used in previous gallium experiments. Therefore, the BEST-2 experiment will also investigate the dependence of the gallium anomaly on the neutrino energy. Observation of a noticeable difference in the total neutrino capture rate in experiments with neutrinos of different energies will mean that sterile oscillations are not the main cause of the gallium anomaly. In this case, it is necessary to look for other causes of the gallium anomaly.

The $^{58}$Co source for the BEST-2 experiment can be produced from 15 kg of natural nickel or from 10 kg of nickel enriched in the $^{58}$Ni isotope, in (n,p) reactions in fast neutron reactors with a flux density of $\Phi \sim 2 \times 10^{15}$ cm$^{-2}$s$^{-1}$.

BEST-2 experiment presents a unique opportunity to detect and study phenomena associated with "new physics".

**Funding.** The work was supported by the Ministry of Science and Higher Education of the Russian Federation within the framework of the program for financing large scientific projects of the national project "Science", grant No. 075-15-2024-541.